\documentclass{article}

\usepackage{nopageno,
times,epsf,amsmath,amssymb,cite}
\usepackage[latin1]{inputenc}
\usepackage[]{caption2}
\usepackage{graphics}
\usepackage{bm}

\newcommand{\nt}{\notag \\  & \times}
\newcommand{\na}{\notag \\  & }

\usepackage{bm}

\begin{document}


\title{Negative-energy and tachyonic solutions
in  the Weinberg-Tucker-Hammer equation for spin 1}

\author{Valeriy V. Dvoeglazov\\
UAF, Universidad de Zacatecas\\
Apartado Postal 636, Suc. 3\\
Zacatecas 98061, Zac., M\'exico\\
Email: valeri@fisica.uaz.edu.mx}
\date{}

\maketitle


\begin{abstract} We considered Weinberg-like equations in the article [1] in order to construct the Feynman-Dyson propagator for the spin-1 particles. 
This construction is based on the concept of the Weinberg field as a system of  four field functions differing  by parity  and  by dual transformations.
%
We also analyzed the recent controversy in the definitions of the Feynman-Dyson propagator for the field operator containing the $S=1/2$ self/anti-self charge conjugate states in the papers by D. Ahluwalia et al~\cite{Ahlu-PR} and by W. Rodrigues Jr. et al~\cite{Rodrigues-PR,Rodrigues-IJTP}. The solution to this mathematical controversy is obvious. I proposed the necessary doubling of  the Fock Space (as in the Barut and Ziino works), thus extending the corresponding Clifford Algebra. However, the logical interrelations of different mathematical foundations with  physical interpretations are not so obvious. 
In this work we present some insights with respect to this for spin 1/2 and 1.

Meanwhile, the N. Debergh et al article considered our old ideas of doubling the Dirac equation, and other forms of T- and PT-conjugation [5]. 
Both algebraic equation  $Det (\hat p - m) =0$ and $Det (\hat p + m) =0$  for  $u-$ and $v-$  4-spinors have solutions with $p_0= \pm E_p =\pm \sqrt{{\bf p}^2 +m^2}$. The same is true for higher-spin equations (or they may even have more complicated dispersion relations). Meanwhile, every book considers 
the equality $p_0=E_p$  for both $u-$  and $v-$ spinors of the $(1/2,0)\oplus (0,1/2))$ representation only, thus applying the Dirac-Feynman-Stueckelberg procedure for elimination of negative-energy solutions. 
The Fock space can be doubled on the quantum-field (QFT) level. 
In this article we give additional bases for the development of the correct theory of higher spin particles in QFT. It seems, that it is imposible to consider  the relativistic quantum mechanics appropriately without  negative energies, tachyons and appropriate forms of the discrete symmetries, and their actions on the corresponding physical states.
\end{abstract}



\large{

\section{Introduction.}

The algebraic equations  $Det (\hat p - m) =0$ and $Det (\hat p + m) =0$  for  $u-$ and $v-$  4-spinors have solutions with $p_0= \pm E_p =\pm \sqrt{{\bf p}^2 +m^2}$. 
The recent problems of superluminal neutrinos, negative-mass squared
neutrinos, various schemes of oscillations including sterile neutrinos,
require  attention. The problem of the lepton mass splitting ($e, \mu ,\tau $) has long
history. This suggests that something missed in the foundations of relativistic
quantum theories. Modifications seem to be necessary in the Dirac sea concept, and in the even more sophisticated 
Stueckelberg concept of the backward propagation in time. The Dirac sea concept is intrinsically related to the Pauli principle. However, the Pauli principle is intrinsically related to the Fermi statistics and the anticommutation relations of fermions. Recently, the concept of the {\it bi-orthonormality} has been proposed; the (anti) commutation relations and statistics are assumed to be different for {\it neutral} particles~\cite{Ahl-new}. We propose the relevant modifications in the basics of the relativistic quantum theory below. 
Next, Sakharov in 1967, Ref.~\cite{Saharov}, introduced the idea of two universes with opposite 
arrows of time, born from the same initial singularity (i.e. Big Bang). Next, the authors of~\cite{Debergh} constructed  (within the framework of 
the present-day quantum field theory) negative-energy fields for spin-$1/2$ fermions. Currently, the predominating consensus is  existence of the dark matter (DM) and the dark energy (DE) paradigm.  Numerous possible candidates have been proposed for the DM, but to the date, search for these candidates was not  successful. ``There is growing favor with the idea that new ideas need to be considered until an answer is found." ``A paradigm shift that allows the serious consideration of negative mass is a real possibility." However, see~\cite{Chub} on the relation of inertial and gravitational masses.

The paper is composed in the following way: Introduction, General Frameworks, Negative-energy and Tachyonic Solutions for the spin 1, and Conclusions. In the main text  the Dirac spinor formalism is given. 

\section{The General Frameworks.}

\subsection{The Dirac formalism with negative energies.}

The Dirac equation is:
\begin{equation}
[i\gamma^\mu \partial_\mu -m]\Psi (x) =0\,.\label{Dirac}
\end{equation}
The $\gamma^\mu$ are the Clifford algebra matrices, ($\mu, \nu = 0,1,2,3$):
\begin{equation}
\gamma^\mu \gamma^\nu +\gamma^\nu \gamma^\mu = 2g^{\mu\nu}\,,
\end{equation}
$g^{\mu\nu}$ is the metrics of the Minkowski space.
Usually, everybody uses the following definition of the field operator~\cite{Ryder,Itzykson} in the pseudo-Euclidean metrics:
\begin{equation}
\Psi (x) = \frac{1}{(2\pi)^3}\sum_h \int \frac{d^3 {\bf p}}{2E_p} [ u_h ({\bf p}) a_h ({\bf p}) e^{-ip\cdot x}
+ v_h ({\bf p}) b_h^\dagger ({\bf p})] e^{+ip\cdot x}]\,,
\end{equation}
as given {\it ab initio}.
After actions of the Dirac operator at  $\exp (\mp ip_\mu x^\mu)$ the 4-spinors ( $u-$ and $v-$ ) 
satisfy the momentum-space equations: $(\hat p - m) u_h (p) =0$ and  $(\hat p + m) v_h (p) =0$, respectively;  $h$ is 
the polarization index. It is easy to prove from the characteristic equations
$Det (\hat p \mp m) =(p_0^2 -{\bf p}^2 -m^2)^2= 0$ that the solutions should satisfy the energy-momentum relation $p_0= \pm E_p =\pm \sqrt{{\bf p}^2 +m^2}$.

The general scheme of construction of the field operator has been presented in~\cite{Bogoliubov}. In the case of
the $(1/2,0)\oplus (0,1/2)$ representation we have:
\begin{eqnarray}
&&\Psi (x) = {1\over (2\pi)^3} \int d^4 p \,\delta (p^2 -m^2) e^{-ip\cdot x}
\Psi (p) =\nonumber\\
&=& {1\over (2\pi)^3} \sum_{h}^{}\int d^4 p \, \delta (p_0^2 -E_p^2) e^{-ip\cdot x}
u_h (p_0, {\bf p}) a_h (p_0, {\bf p}) =\label{fo}\\
&=&{1\over (2\pi)^3} \int {d^4 p \over 2E_p} [\delta (p_0 -E_p) +\delta (p_0 +E_p) ] \nonumber\\
&&[\theta (p_0) +\theta (-p_0) ] e^{-ip\cdot x}
\sum_{h}^{} u_h (p) a_h (p) =\nonumber\\
&=& {1\over (2\pi)^3} \sum_h^{} \int {d^4 p \over 2E_p} [ \delta (p_0 -E_p) +\delta (p_0 +E _p) ] 
\nonumber\\
&&\left
[ \theta (p_0) u_h (p) a_h (p) e^{-ip\cdot x}  + 
\theta (p_0) u_h (-p) a_h (-p) e^{+ip\cdot x} \right ] \nonumber\\
&=& {1\over (2\pi)^3} \sum_h^{} \int {d^3 {\bf p} \over 2E_p} \theta(p_0)\nonumber\\  
&&\left [ u_h (p) a_h (p)\vert_{p_0=E_p} e^{-i(E_p t-{\bf p}\cdot {\bf x})}  +
u_h (-p) a_h (-p)\vert_{p_0=E_p} e^{+i (E_p t- {\bf p}\cdot {\bf x})}
\right ],\nonumber
\end{eqnarray}
where $a_h, b_h^\dagger$ are the annihilation/creation operators, and in the textbook cases
\begin{eqnarray}
u_h ({\bf p}) = 
\begin{pmatrix}
\exp ( + {\bm \sigma } \cdot {\bm \varphi }/2 ) \phi_R^h ( {\bf 0 } )\cr
\exp ( - {\bm \sigma }\cdot {\bm \varphi }/2 )\phi_L^h ( {\bf 0} )  \cr 
\end{pmatrix} \,
\end{eqnarray}
$cosh (\varphi)=E_p/m$, $sinh (\varphi) =\vert {\bf p} \vert /m $.
The 2-spinors are $\phi_L^{\uparrow\downarrow} ({\bf 0})=\begin{pmatrix} 1\cr 0\cr \end{pmatrix}, \begin{pmatrix} 
0\cr 1\cr\end{pmatrix}$, $\phi_R^h =\mp \Theta_{[1/2]}\phi_L^{-h}$, $\Theta_{[1/2]} =\begin{pmatrix}0&-1\cr 1&0\cr
\end{pmatrix}$.
During the calculations above we had to represent $1=\theta (p_0) +\theta (-p_0)$
in order to get positive- and negative-frequency parts~\cite{DvoeglazovJPCS}.
Moreover, during these calculations we did not yet assume, which equation this
field operator  (namely, the $u-$ spinor) satisfies, with negative- or positive- mass?\footnote{Moreover, since bispinors are, in general, complex-valued,
we can even use the different basis such as $u_\alpha = column (i\,0\,0\,0)$ etc. instead of the well-accustomed one.} 
In general, we should transform $u_h (-p)$ to $v_h (p)$. The procedure is the following one~\cite{DvoeglazovHJ}.
In the Dirac case we should assume the following relation in the field operator:
\begin{equation}
\sum_{h}^{} v_h (p) b_h^\dagger (p) = \sum_{h}^{} u_h (-p) a_h (-p)\,.\label{dcop}
\end{equation}
We need $\Lambda_{\mu\lambda} (p) = \bar v_\mu (p) u_\lambda (-p)$.
By direct calculations,  we find
\begin{equation}
-mb_\mu^\dagger (p) = \sum_{\lambda}^{} \Lambda_{\mu\lambda} (p) a_\lambda (-p)\,.
\end{equation}
Hence, $\Lambda_{\mu\lambda} = -im ({\bm \sigma}\cdot {\bf n})_{\mu\lambda}$, ${\bf n} = {\bf p}/\vert{\bf p}\vert$, 
and 
\begin{equation}
b_\mu^\dagger (p) = i\sum_\lambda ({\bm\sigma}\cdot {\bf n})_{\mu\lambda} a_\lambda (-p)\,.
\end{equation}
Multiplying (\ref{dcop}) by $\bar u_\mu (-p)$ we obtain
\begin{equation}
a_\mu (-p) = -i \sum_{\lambda} ({\bm \sigma} \cdot {\bf n})_{\mu\lambda} b_\lambda^\dagger (p)\,.
\end{equation}
The equations are self-consistent. In the $(1,0)\oplus (0,1)$ representation 
the similar procedure leads to the different situation:
\begin{equation}
a_\mu (p) = [1-2({\bf S}\cdot {\bf n})^2]_{\mu\lambda} a_\lambda (-p)\,. 
\end{equation}
This signifies that in order to construct the Sankaranarayanan-Good field operator~\cite{Sankar}, it satisfies 
$[\gamma_{\mu\nu} \partial_\mu \partial_\nu - {(i\partial/\partial t)\over E} 
m^2 ] \Psi (x) =0$, we need additional postulates. For instance, one can try to construct the left- and the right-hand side of the field operator separately each other~\cite{DvoeglazovJPCS}.

We have, in fact,
$u_h ( E_p, {\bf p} )$ and $u_h (-E_p, {\bf p})$ originally, 
which satisfy the equations:
\begin{equation}
\left [ E_p (\pm \gamma^0) - {\bm \gamma}\cdot {\bf p} - m \right ] u_h (\pm E_p, {\bf p})=0\,.
\end{equation}
Due to the properties $ U^\dagger \gamma^0 U=-\gamma^0$, $U^\dagger \gamma^i U=+\gamma^i$
with the unitary matrix
\begin{eqnarray}
U= \begin{pmatrix} 0 &-1\cr 1 & 0\cr  \end{pmatrix}
=  \gamma^0\gamma^5 
\end{eqnarray}
in the Weyl basis,\footnote{The properties of the $U-$ matrix are opposite to those of
$P^\dagger \gamma^0 P=+\gamma^0$, $P^\dagger \gamma^i P=-\gamma^i$
with the usual $P=\gamma^0$, thus giving $\left [ -E_p \gamma^0 + {\bf \gamma}\cdot {\bf p} - m \right ] 
P u_h (- E_p, {\bf p}) = -\left [\hat p +m \right ] \tilde v_{?} (p) = 0$. While, the relations of the spinors 
$v_h (E_p, {\bf p})=\gamma_5  u_h (E_p, {\bf p})$ are well-known, it seems that
the relations of the $v-$ spinors of the positive energy to $u-$ spinors of the negative energy
are frequently forgotten, $\tilde v_{?} (p) = \gamma^0 u_h (- E_p, {\bf p})$. } 
we have in the negative-energy case:
\begin{equation}
\left [ E_p \gamma^0 - {\bm \gamma}\cdot {\bf p} - m \right ] U^\dagger u_h (- E_p, {\bf p})=0\,.\label{nede}
\end{equation}
Thus, unless the unitary transformations do not change
the physical content, we have that the negative-energy spinors $\gamma^5 \gamma^0 u (- E_p, {\bf p})$ (see (\ref{nede})) satisfy the accustomed ``positive-energy" Dirac equation. Their explicite forms $\gamma^5 \gamma^0 u (- E_p, {\bf p})$ are different from the textbook ``positive-energy" Dirac spinors.
From the first sight (just  $E_p \rightarrow - E_p$) they are the following ones:
\begin{eqnarray}
\tilde u (p) = \frac{N}{\sqrt{2m (-E_p +m)}} \begin{pmatrix}-p^+ + m\cr -p_r\cr
p^- -m \cr - p_r\cr\end{pmatrix}\,,\\
\tilde{\tilde u} (p) =\frac{N}{\sqrt{2m (-E_p +m)}}\begin{pmatrix}-p_l \cr -p^- + m\cr
-p_l \cr p^+ -m\cr\end{pmatrix}\,.
\end{eqnarray}
We use tildes because we do not yet know their polarization properties. It is not even clear, which helicity operator,
$\sigma_3/2$ or $({\bm \sigma} \cdot  {\bf n})/2$, or some other  should be used after $T-$ conjugation~\cite{Debergh}. Next,
\begin{equation}
E_p=\sqrt{{\bf p}^2 +m^2}>0, p_0=\pm E_p, p^\pm = E_p\pm p_z, p_{r,l}= p_x\pm ip_y.
\end{equation}
What about the $\tilde v (p)=\gamma^0 u$ transformed with the $\gamma_0$ matrix?
They are not equal to the previous ``negative-energy"  4-spinors $ v_h (p) =\gamma^5 u_h (p)$?
Obviously, they also do not have well-known forms  of the usual $v-$ spinors in the Weyl basis, 
differing by phase factor and in the sign at the mass term. The normalizations of these 4-spinors are to $(\pm 2N^2)$.

One can prove that the matrix
\begin{equation}
P= e^{i\theta}\gamma^0 = e^{i\theta}\begin{pmatrix}0& 1_{2\times 2}\cr 1_{2\times 2} & 0\cr\end{pmatrix}
\label{par}
\end{equation}
can be used in the parity operator as well as
in the original Weyl basis. 
However, if we would take the phase factor to be zero
we obtain that while $u_h (p)$ have the eigenvalue $+1$, but after space inversion operation
($R= ({\bf x} \rightarrow -{\bf x}, {\bf p} 
\rightarrow -{\bf p}$))
\begin{eqnarray}
P R\tilde u (p) &=& P R\gamma^5 \gamma^0  u_\uparrow (-E_p, {\bf p})= -\tilde u (p)\,,\,\,\\ 
P R \tilde{\tilde u} (p) &=& P R \gamma^5 \gamma^0  u_\downarrow (-E_p, {\bf p})= -\tilde{\tilde u} (p)\,.
\end{eqnarray}
Perhaps, one should choose the phase factor $\theta=\pi$. Thus, we again confirmed
that the relative (particle-antiparticle) intrinsic parity has physical significance only.

Similar formulations have been  presented in Refs.~\cite{Markov}, 
and~\cite{BarutZiino}. The group-theoretical basis for such doubling has been given
in the papers by Gelfand, Tsetlin and Sokolik~\cite{Gelfand}, who first presented 
the theory in the 2-dimensional representation of the inversion group in 1956 (later called as ``the Bargmann-Wightman-Wigner-type quantum field theory" in 1993). M. Markov wrote long ago {\it two} Dirac equations with  the opposite signs at the mass term~\cite{Markov}. 
\begin{eqnarray}
\left [ i\gamma^\mu \partial_\mu - m \right ]\Psi_1 (x) &=& 0\,,\\
\left [ i\gamma^\mu \partial_\mu + m \right ]\Psi_2 (x) &=& 0\,.
\end{eqnarray}
In fact, he studied all properties of this relativistic quantum model (while he did not know yet the quantum
field theory in 1937). Next, he added and  subtracted these equations:
\begin{eqnarray}
i\gamma^\mu \partial_\mu \varphi (x) - m \chi (x) &=& 0\,, \label{sas1}\\
i\gamma^\mu \partial_\mu \chi (x) - m \varphi (x) &=& 0\,. \label{sas2}
\end{eqnarray}
Thus, $\varphi$ and $\chi$ solutions can be presented as some superpositions of the Dirac 4-spinors $u-$ and $v-$.
These equations, of course, can be identified with the equations for the Majorana-like $\lambda -$ and $\rho -$, which we presented 
in Ref.~\cite{Dvoegl-NP} on the basis of postulates~\cite{Ahlu-NP}. 



The  four-component Majorana-like spinors are
	\begin{align}
	\lambda({\bf p})= \left(\begin{array}{c}
 	\vartheta\Theta \phi^\ast_L({\bf p})\\
	\phi_L ({\bf p})
	\end{array}\right)\,,\label{eq:taup}
	\end{align}
see $\Theta_{[1/2]}$ above.
They become eigenspinors of the charge conjugation operator $S_c$ with eigenvalues $\pm 1$ if the phase $\vartheta$ is set to  $\pm \, i$:
	\begin{align}
		S_c\; \lambda({\bf p})\Big\vert_{\vartheta=\pm i}
                = \pm \lambda({\bf p})\Big\vert_{\vartheta=\pm i}. \label{eq:tau}
	\end{align}
In a similar way one can construct $\rho-$ spinors on using $\phi_R$.

Of course, the signs at the mass terms
depend on, how do we associate the positive- or negative- frequency solutions with $\lambda$ and $\rho$:
\begin{eqnarray}
i \gamma^\mu \partial_\mu \lambda^S (x) - m \rho^A (x) &=& 0 \,,
\label{11}\\
i \gamma^\mu \partial_\mu \rho^A (x) - m \lambda^S (x) &=& 0 \,,
\label{12}\\
i \gamma^\mu \partial_\mu \lambda^A (x) + m \rho^S (x) &=& 0\,,
\label{13}\\
i \gamma^\mu \partial_\mu \rho^S (x) + m \lambda^A (x) &=& 0\,,
\label{14}
\end{eqnarray}
$S$ and $A$ are self- and anti-self charge conjugate states.
Neither of them can be regarded as the Dirac equation.
However, they can be written in the 8-component form as follows:
\begin{eqnarray}
\left [i \Gamma^\mu \partial_\mu - m\right ] \Psi_{_{(+)}} (x) &=& 0\,,
\label{psi1}\\
\left [i \Gamma^\mu \partial_\mu + m\right ] \Psi_{_{(-)}} (x) &=& 0\,,
\label{psi2}
\end{eqnarray}
with
\begin{eqnarray}
&&\hspace{-20mm}\Psi_{(+)} (x) = \begin{pmatrix}\rho^A (x)\cr
\lambda^S (x)\cr\end{pmatrix},
\Psi_{(-)} (x) = \begin{pmatrix}\rho^S (x)\cr
\lambda^A (x)\cr\end{pmatrix}, \quad\mbox{and}\quad\Gamma^\mu =\begin{pmatrix}0 & \gamma^\mu\cr
\gamma^\mu & 0\cr\end{pmatrix}.
\end{eqnarray}
It is easy to find the corresponding projection operators, and the Feynman-Stueckelberg propagator.

The connection with the Dirac spinors has 
been found~\cite{Dvoegl-NP,Kirchbach}.
For instance,
\begin{eqnarray}
\begin{pmatrix}\lambda^S_\uparrow ({\bf p}) \cr \lambda^S_\downarrow ({\bf p}) \cr
\lambda^A_\uparrow ({\bf p}) \cr \lambda^A_\downarrow ({\bf p})\cr\end{pmatrix} = {1\over
2} \begin{pmatrix}1 & i & -1 & i\cr -i & 1 & -i & -1\cr 1 & -i & -1 & -i\cr i&
1& i& -1\cr\end{pmatrix} \begin{pmatrix}u_{+1/2} ({\bf p}) \cr u_{-1/2} ({\bf p}) \cr
v_{+1/2} ({\bf p}) \cr v_{-1/2} ({\bf p})\cr\end{pmatrix},\label{connect}
\end{eqnarray}
provided that the 4-spinors have the same physical dimension.
Thus, we can see
that the two 4-spinor systems are connected by  unitary transformations, and this represents
itself rotations of the spin-parity basis. However, it is usually assumed that the $\lambda-$ and $\rho-$ spinors describe the neutral particles,
meanwhile $u-$ and $v-$ spinors describe the charged particles. Kirchbach {\it et al.}~\cite{Kirchbach} found the amplitudes for 
neutrinoless double beta decay ($00\nu\beta$) in this scheme. It is obvious from (\ref{connect}) that there are some additional terms in $00\nu\beta$ comparing with the standard formulation.  
One can also re-write the above equations into the two-component forms. Thus, one obtains the Feynman-Gell-Mann~\cite{FG} equations.

Barut and Ziino~\cite{BarutZiino} proposed yet another model. They considered
$\gamma^5$ operator as the operator of the charge conjugation. Thus, the charge-conjugated
Dirac equation has the different sign comparing with the ordinary formulation:
\begin{equation}
[i\gamma^\mu \partial_\mu + m] \Psi_{BZ}^c =0\,,
\end{equation}
and the so-defined charge conjugation applies to the whole system, \linebreak 
fermion+electro\-magnetic field, $e\rightarrow -e$
in the covariant derivative. The superpositions of the $\Psi_{BZ}$ and $\Psi_{BZ}^c$ also give us 
the ``doubled Dirac equation", as the equations for $\lambda-$ and $\rho-$ spinors. 
The concept of the doublings of the Fock space has been
developed in the Ziino works (cf.~\cite{Gelfand,DvoeglazovBW}) in the framework of the quantum field theory. In their case the charge conjugate states
are simultaneously the eigenstates of the chirality.
Next, it is interesting to note that for the Majorana-like field operator  ($\nu$ refers to neutrino)
we have
\begin{equation}
\nu^{^{ML}} (x^\mu) = 
\int {d^3{\bf p} \over (2\,\pi)^3  } {1\over {2E_p}}
\sum_{\eta}
\left[
\lambda^S_{\eta}(p^\mu)\,a_{\eta}(p^\mu)\, e^{-\,i\,p\cdot  x}
\,+\,
\lambda^A_{\eta}(p^\mu)\,a^\dagger_{\eta}(p^\mu)\,
e^{+\,i\,p\cdot x}\right]
\,.
\end{equation}
Hence,
\begin{eqnarray}
&&\left [ \nu^{^{ML}} (x^\mu) + {\cal C} \nu^{^{ML\,\dagger}} (x^\mu) \right
]/2 =\nonumber\\
&=& \int {d^3 {\bf p} \over (2\pi)^3 } {1 \over 2E_p} \sum_\eta \left [ 
\begin{pmatrix}
i\Theta \phi_{_L}^{\ast \, \eta} (p^\mu) \cr 0 \end{pmatrix} a_\eta
(p^\mu)  e^{-ip\cdot x} +\right.\nonumber\\
&+&\left. \begin{pmatrix} 0\cr
\phi_L^\eta (p^\mu)\cr
\end{pmatrix}
a_\eta^\dagger (p^\mu) e^{ip\cdot x} \right ]\,\nonumber ,\\
&&\left [\nu^{^{ML}} (x^\mu) - {\cal C} \nu^{^{ML\,\dagger}} (x^\mu) \right
]/2 =\\
&=&\int {d^3 {\bf p} \over (2\pi)^3 } {1\over 2E_p} \sum_\eta \left [
\begin{pmatrix}0\cr \phi_{_L}^\eta (p^\mu) \end{pmatrix} a_\eta (p^\mu)  e^{-ip\cdot x}
+\right.\nonumber\\
&+&\left.\begin{pmatrix}-i\Theta \phi_{_L}^{\ast}\,{ \eta} (p^\mu)\cr 0\end{pmatrix} 
a_\eta^\dagger (p^\mu) e^{ip\cdot x} \right ]\, , \nonumber
\end{eqnarray}
which, thus, naturally lead to the Ziino-Barut scheme of massive chiral
fields, Ref.~\cite{BarutZiino}.
Next, the relevant paper is Ref.~\cite{Fabbri}. It is obvious to merge $u ({\bf p})$ and $v ({\bf p})$ spinors in one doublet of ``positive energy" and $v ({\bf p})$ and $u ({\bf p})$ spinors, in another doublet of ``negative energy" , as Markov and Fabbri did. 
However, the point of my paper is that both $u (p_0, {\bf p})$ and $v (p_0, {\bf p})$ contains contributions to both positive- and negative- energies, cf.~\cite{Ni}.

\subsection{The Feynman-Dyson propagators for neutral particles (locality or non-locality?).}


We study the problem of construction of causal propagators in both higher-spin theories and the spin $S=1/2$
Majorana-like theory. The hypothesis is: in order to construct the analogues of the Feynman-Dyson propagator 
we need actually four field operators connected by the dual and parity transformation. We use the standard methods of quantum field theory.
So, the number of components in the causal propagators is enlarged accordingly. The conclusion is discussed in the last Section: if we 
would not enlarge the number of components in the fields (in the propagator) we would not be able to obtain the causal 
propagator.

Accordingly  to the Feynman-Dyson-Stueckelberg ideas,
a causal propagator  $S_{F}$ has to be constructed
on using  the formula (\textit{e.g}., Ref.~\cite[p.91]{Itzykson})
\begin{align}
S_F (x_2, x_1) & =\sum_\sigma\int \frac{d^3 p}{(2\pi)^3}\frac{m}{E_p}
\nt
\big[
\theta (t_2 -t_1) \, a \,\,
u^\sigma (p)  \overline u^{\sigma} (p) e^{-ip\cdot x} 
\na
+  
\theta (t_1 - t_2) \, b \,\,
v^\sigma (p)  \overline v^\sigma (p) e^{ip\cdot x} 
\big] \, ,
\end{align}
where $x=x_2 -x_1$, $m$ is the particle mass, 
$p^\mu = (E_p, {\bf p} )$, $u^\sigma$, $v^\sigma$ 
are the 4-spinors, $\theta (t)$ is the Heaviside
function. In the spin $S=1/2$ Dirac theory,  it results  in
\begin{align}\label{dp1}
S_F (x) = \int \frac{d^4 p}{(2\pi)^4} e^{-ip\cdot x} \frac{\hat p +m}{p^2 -m^2
+i\epsilon} \,,
\end{align}
provided that the constants $a$ and $b$ are determined by imposing
\begin{align}
(i\hat \partial_2 -m) S_F (x_2, x_1) =\delta^{(4)} (x_2 -x_1) \, ,
\end{align}
namely, $a=-b=1/ i$; $\partial_2 =\partial/\partial x_2$, $\epsilon$ defines the rules of work near the poles.

However,  attempts to construct the covariant propagator in this way
have failed in the framework of the Weinberg theory,
Ref.~\cite{Weinberg}, which is a generalization of the Dirac ideas to
higher spins. For instance, on the page B1324 of Ref.~\cite{Weinberg}\,
Weinberg writes:

``{\it Unfortunately,
the propagator arising from Wick's theorem is  NOT equal to the covariant
propagator
except for $S=0$ and $S=1/2$. The trouble is that the derivatives act on the
$\epsilon (x) = \theta (x) - \theta (-x)$ in $\Delta^C (x)$ as well as on
the functions\footnote{In the cited paper $\Delta_1(x) \equiv i \left [\Delta_+
(x) + \Delta_+ (-x)\right ]$ and $\Delta (x) \equiv \Delta_+ (x) - \Delta_+
(-x)$ have been used.
 $i\Delta_+ (x) \equiv ({1}/{(2\pi)^3}) \int ({d^3 p}/{2E_p}) \exp
(ip\cdot x)$ is the particle Green function.} $\Delta$ and $\Delta_1$. This
gives rise to extra terms proportional to equal-time $\delta$ functions
and their derivatives\ldots The cure is well known: \ldots compute the
vertex factors using only the original covariant part of the Hamiltonian
${\cal H}$; do not use the Wick propagator for internal lines; instead use
the covariant propagator.}

The propagator proposed  in Ref.~\cite{Ah-pro} is the causal propagator.
However, the old problem remains: the Feynman-Dyson propagator
is not the Green function of the Weinberg equation. As mentioned,
the covariant propagator proposed by Weinberg propagates kinematically
spurious solutions~\cite{Ah-pro}.

The aim of this subsection is to consider
the problem of constructing the propagator in
the framework of the model given in~\cite{D1}.
The concept of
the Weinberg field {\it doubles}  has been proposed there.
It is based on the equivalence between
the Weinberg field and the  antisymmetric tensor field, 
which can be described by both $F_{\mu\nu}$ and
its dual $\tilde F_{\mu\nu}$.
These field functions may be used to form a parity doublet. An essential
ingredient of my consideration
is the idea of combining the Lorentz and the dual
transformation. 
For the functions $\psi_1^{(1)}$ and $\psi_2^{(1)}$, connected with the
first one by  the dual (chiral, $\gamma_5 =diag (1_{3\times 3}), -1_{3\times 3})$)  transformation,
the equations are\footnote{I have to use the Euclidean metrics here in order a reader to be able to compare
the formalism with the classical cited works. In the next Sections we turn to the pseudo-Euclidean metrics on using simple correspondence rules.}
\begin{align}\label{eq:a1}
(\gamma_{\mu\nu} p_\mu p_\nu +m^2 )\psi_1^{(1)} =& 0\,,\\ \label{eq:a2}
(\gamma_{\mu\nu} p_\mu p_\nu - m^2) \psi_2^{(1)} =& 0 \,,
\end{align}
with $\mu,\nu =1,2,3,4$. For the field functions
connected with $\psi_1^{(1)}$ and $\psi_2^{(1)}$ by the $\gamma_5\gamma_{44}$
transformations the set of equations is written:
\begin{align}\label{eq:a11}
\left [\widetilde \gamma_{\mu\nu}p_\mu p_\nu - m^2\right ] \psi_1^{(2)} =&0
\,,\\
\label{eq:a21}
\left [\widetilde \gamma_{\mu\nu} p_\mu p_\nu + m^2 \right ] \psi_2^{(2)} =&0
\,,
\end{align}
where $\widetilde \gamma_{\mu\nu} = \gamma_{44} \gamma_{\mu\nu} \gamma_{44}$
is connected with the $S=1$ Barut-Muzinich-Williams $\gamma_{\mu\nu}$ matrices~\cite{Barut, Hammer}.
In the cited paper  I  have used the plane-wave expansion.
The corresponding `bispinors'
in the momentum space coincide with the Tucker-Hammer ones within
a normalization.\footnote{They  also coincide with the Ahluwalia
{\it et al.} ones within a unitary transformation~\cite{Ahlu-PR}.}   Their
explicit forms are 
\begin{eqnarray}\label{b1}
u_1^{\sigma\, (1)} ({\bf p}) = v_1^{\sigma\, (1)} ({\bf p})
=\frac{1}{\sqrt{2}}
\begin{pmatrix}
\left [  
m+ ({\bf S}\cdot{\bf p})
+{({\bf S} \cdot {\bf p})^2 \over  (E+m)}
\right ]
\xi_\sigma \\
 \left [ 
m  -
({\bf S}\cdot{\bf p}) +{({\bf S}\cdot{\bf p})^2 \over  (E+m)}
\right ]
\xi_\sigma 
\end{pmatrix}
\end{eqnarray}
and
\begin{eqnarray}\label{b12}
u_2^{\sigma\, (1)} ({\bf p}) = v_2^{\sigma\, (1)} ({\bf p})
=\frac{1}{\sqrt{2}}
\begin{pmatrix}
\left [  
m+ ({\bf S}\cdot{\bf p})
+{({\bf S} \cdot {\bf p})^2 \over  (E+m)}
\right ]
\xi_\sigma \\
 \left [ 
- m + ({\bf S}\cdot{\bf p}) -{({\bf S}\cdot{\bf p})^2 \over  (E+m)}
\right ]
\xi_\sigma 
\end{pmatrix}
\end{eqnarray}
where $\xi_\sigma$ are the 3-component objects (the analogs of the Weyl spinors).
Thus,  $u_2^{(1)} ({\bf p}) = \gamma_5 u_1^{(1)} ({\bf p})$ and
$\overline u_2^{(1)} ({\bf p}) = -\overline u_1^{(1)} ({\bf p})\gamma_5$.
The bispinors
\begin {eqnarray}
\label{b11}
&&u_1^{\sigma\, (2)} ({\bf p}) = v_1^{\sigma\, (2)}
({\bf p}) 
=\frac{1}{\sqrt{2}}
\begin{pmatrix}
\left [ m- ({\bf S}\cdot {\bf p})+{{\bf S}\cdot {\bf p})^2 \over  (E+m)}\right ]\xi_\sigma \\
\left [ -m  - ({\bf S}\cdot {\bf p}) -{({\bf S}\cdot {\bf p})^2 \over  (E+m)}\right ] \xi_\sigma 
\end{pmatrix}
\\
&&u_2^{\sigma\,(2)} ({\bf p})  = v_2^{\sigma\,(2)} ({\bf p})
=\frac{1}{\sqrt{2}}
\begin{pmatrix}
\left [ -m+ ({\bf S}\cdot{\bf p}) -{({\bf S}\cdot {\bf p})^2 \over  (E+m)}\right ]\xi_\sigma  \\
 \left [  -m  - ({\bf S}\cdot{\bf p}) - {({\bf S}\cdot {\bf p})^2 \over  (E+ m)}\right ] \xi_\sigma
\end{pmatrix}
\end{eqnarray}
\noindent 
satisfy Eqs. (\ref{eq:a11}) and (\ref{eq:a21}) written in the momentum space.
Thus,
$u_1^{(2)} ({\bf p}) = \gamma_5\gamma_{44} u_1^{(1)} ({\bf p})$,
$\overline u_1^{(2)} = \overline u_1^{(1)} \gamma_5\gamma_{44}$,
$u_2^{(2)} ({\bf p}) = \gamma_5\gamma_{44} \gamma_5 u_1^{(1)} ({\bf p})$
and $\overline u_2^{(2)} ({\bf p}) = - \overline u_1^{(1)}\gamma_{44}$.

Let me check, if the sum of four equations
\begin{eqnarray}
&& \hspace{-25mm}\left [ \gamma_{\mu\nu} \partial_\mu \partial_\nu  - m^2 \right ]
 \int  \frac{d^3 p}{(2\pi)^3 2E_p}
\left [ \theta (t_2 -t_1) \, a\,\,\,   u_1^{\sigma\,(1)} (p)   \overline
u_1^{\sigma\,(1)} (p) e^{ip\cdot x}
+ \theta (t_1 -t_2) \, b \,\,\, v_1^{\sigma\,(1)} (p)
  \overline  v_1^{\sigma\,(1)} (p) e^{-ip\cdot x} \right  ]+ \nonumber\\
&&\hspace{-25mm}\left [ \gamma_{\mu\nu} \partial_\mu \partial_\nu + m^2 \right  ]  \int
\frac{d^3 p}{(2\pi)^3 2E_p}
\left [ \theta (t_2 -t_1) \, a\,\,\,   u_2^{\sigma\,(1)} (p)   \overline
u_2^{\sigma\,(1)} (p) e^{ip\cdot x}
+\theta (t_1 -t_2) \, b \,\,\, v_2^{\sigma\,(1)} (p)
  \overline  v_2^{\sigma\,(1)} (p) e^{-ip\cdot x}\right  ]+ \nonumber\\
&&\hspace{-25mm}\left [ \widetilde \gamma_{\mu\nu} \partial_\mu \partial_\nu + m^2 \right  ]
\int \frac{d^3 p}{(2\pi)^3 2E_p}
\left [ \theta (t_2 -t_1) \, a\,\,\,  u_1^{\sigma\,(2)} (p)   \overline
u_1^{\sigma\,(2)} (p) e^{ip\cdot x}
+\theta (t_1 -t_2) \, b \,\,\, v_1^{\sigma\,(2)} (p)
  \overline  v_1^{\sigma\,(2)} (p)e^{-ip\cdot x} \right ] +\nonumber\\
&&\hspace{-25mm}\left [\widetilde \gamma_{\mu\nu} \partial_\mu \partial_\nu - m^2 \right  ] \int
\frac{d^3 p}{(2\pi)^3 2E_p}
\left [ \theta (t_2 -t_1) \, a\,\,\,   u_2^{\sigma\,(2)} (p)   \overline
u_2^{\sigma\,(2)} (p)  e^{ip\cdot x}+\theta (t_1 -t_2) \, b  v_2^{\sigma\,(2)} (p)
  \overline  v_2^{\sigma\,(2)} (p) e^{-i\cdot px} \right ]\nonumber\\
&=& 
\delta^{(4)} (x_2 -x_1)
\end{eqnarray}
can be satisfied by the definite choice of $a$ and $b$.
The relation  $ u_i (p) =  v_i (p)$ for bispinors in the momentum space
had been used in Ref.~\cite{D1}.  In the process of
calculations  I  assume
that the 3-``spinors" are normalized to $\delta_{\sigma\sigma^\prime}$\, .
The simple calculations give
\begin{eqnarray}
&&\partial_\mu \partial_\nu  \left[ a\, \theta (t_2 -t_1)\, e^{ip(x_2 -x_1)} 
+ b\, \theta (t_1 -t_2)\, e^{-ip(x_2 -x_1)}  \right ]
=\nonumber\\
&& -  \big [ 
a\, p_\mu p_\nu \theta (t_2 - t_1)
\exp \left [ ip(x_2 -x_1)\right ] 
+
b\,  p_\mu p_\nu  \theta (t_1 -t_2)
\exp \left [ -ip (x_2 -x_1) \right ] \big ] \nonumber\\
&&+ a \big [ - \delta_{\mu 4} \delta_{\nu 4} \delta^{\,\,\prime}
(t_2 -t_1) 
+i (p_\mu \delta_{\nu 4} +p_\nu \delta_{\mu 4}) \delta (t_2 -t_1)
\big ] \nonumber\\
&&\exp \left [ i p \cdot ( x_2 - x_1 )\right ] 
+ b\, \left [ \delta_{\mu 4} \delta_{\nu 4} \delta^{\,\,\prime}
(t_2 -t_1) + \right .\nonumber\\
&&\left . i ( p_\mu \delta_{\nu 4} +p_\nu \delta_{\mu 4})
\delta ( t_2 -t_1 ) \right ] \exp \left [ -ip(  x_2 - x_1)\right ]\, ;
\end{eqnarray}
\noindent 
and
\begin{eqnarray}
&&u_1^{(1)}\overline u_1^{(1)}  ={1\over 2} 
\left(
\begin{array}{*{20}c}
m^2 & S_p \otimes S_p \\
\overline S_p \otimes \overline S_p &m^2 \\
\end{array}
\right),\\
&&u_2^{(1)}\overline u_2^{(1)} = {1\over 2}
\left(
\begin{array}{*{20}c}
-m^2  & S_p \otimes S_p  \\
 \overline S_p \otimes \overline S_p &-m^2 \\
\end{array}
\right ),
\end{eqnarray}
\begin{eqnarray}
&&u_1^{(2)}\overline u_1^{(2)}={1\over 2} 
\left(
\begin{array}{*{20}c}
-m^2 & \overline S_p \otimes \overline  S_p\\
S_p \otimes  S_p &-m^2 \\
\end{array}
\right ),\\
&&u_2^{(2)}\overline u_2^{(2)} = {1\over 2}
 \left(
\begin{array}{*{20}c}
m^2  \overline S_p \otimes \overline S_p \\
S_p \otimes  S_p &m^2  \\
\end{array}
\right ) ,
\end{eqnarray}
where
\begin{eqnarray}
S_p &=& m + ({\bf S} \cdot{\bf p}) +\frac{({\bf S} \cdot{\bf p})^2}{E+m}\,,\\
\overline S_p &=& m - ({\bf S} \cdot {\bf p}) + \frac{({\bf S} \cdot {\bf p})^2}{E+m}\,
\end{eqnarray}
are the Lorentz boost matrices.
Due to
\begin{eqnarray}
&&\left [E_p - ({\bf S}\cdot {\bf p})\right ]  S_p \otimes S_p = m^2 \left [ E_p
+ ({\bf S}\cdot {\bf p})\right ]\, ,\\
&&\left [E_p + ({\bf S}\cdot {\bf p})\right ] \overline S_p
\otimes \overline S_p = m^2 \left [ E_p - ({\bf S}\cdot {\bf p})\right ]\,  ,
\end{eqnarray}
one  can conclude: the  generalization of
the notion of causal  propagators is  admitted by using the
Wick-like formula for the time-ordered particle operators
provided that  $a=b=1/ 4im^2$. It is necessary to  consider
all four equations, Eqs. (\ref{eq:a1})-(\ref{eq:a21}). Obviously, this is related to the 12-component formalism, which I presented 
in~\cite{D1}.

The $S=1$ analogues of the formula (\ref{dp1})  for the Weinberg propagators
follow immediately. In the Euclidean metrics they are:
\begin{align}\label{propa1}
S_F^{(1)} ( p )  &{\sim} -\frac{1}{i(2\pi)^4  (p^2  {+}m^2 {-}i\epsilon)} \left [
\gamma_{\mu\nu} p_\mu p_\nu  {-}  m^2  \right ],
\\
\label{propa2}
S_F^{(2)} ( p )& {\sim} -\frac{1}{i(2\pi)^4  (p^2  {+}m^2 {-} i\epsilon)} \left [
\gamma_{\mu\nu} p_\mu p_\nu   {+}  m^2  \right ],
\\
\label{propa3}
S_F^{(3)} ( p ) &{\sim} -\frac{1}{i(2\pi)^4  (p^2  {+} m^2 -i\epsilon)} \left [
\widetilde\gamma_{\mu\nu} p_\mu p_\nu   {+}  m^2  \right ],
\\
\label{propa4}
S_F^{(4)} ( p ) &{\sim} -\frac{1}{i(2\pi)^4  (p^2  {+}m^2 {-}i\epsilon)} \left [
\widetilde \gamma_{\mu\nu} p_\mu p_\nu   {-}  m^2  \right ] .
\end{align}

The controversy in the case of $\lambda-$ and $\rho-$ spinors of the $(1/2,0)\oplus (0,1/2)$ representation is: 
I cited Ahluwalia et al., Ref.~\cite{Ahlu-PR}:\footnote{The notation should be compared with the cited papers.}
``{\it To study the locality structure of the fields $\Lambda(x)$ and $\lambda(x)$, we observe that field momenta  are
	\begin{align}
		\Pi(x)= \frac{\partial{\mathcal L}^\Lambda}
		{\partial\dot\Lambda} =
		\frac{\partial}{\partial t}\stackrel{\neg}{\Lambda}(x),\quad
\end{align}
 and similarly $ \pi(x) = ({\partial}/{\partial
 t})\stackrel{\neg}{\lambda}(x)$.  The calculational details for the
 two fields now differ significantly. We begin with the evaluation of
 the equal time anticommutator for  $\Lambda(x)$ and its conjugate
 momentum
	\begin{align} 
	&\{\Lambda({\bf x},t),\; \Pi({\bf x}^\prime,t)\} =
	i\int\frac{d^3 p}{(2\pi)^3}\frac{1}{2 m} e^{i {\mathbf p}\cdot
	({\mathbf x}-{\mathbf x}^\prime)}
\notag \\ & \hspace{30pt} \times
\underbrace{\sum_\alpha\left[ \xi_\alpha({\bf p})
	\stackrel{\neg}{\xi}_\alpha({\bf p}) - \zeta_\alpha(- {\bf p})
	\stackrel{\neg}{\zeta}_\alpha(- {\bf p})\right]}_{=\, 2 m [ I +
	{\mathcal G}(\mathbf{p})]} .\nonumber 
	\end{align}
The term containing
 ${\mathcal G}({\bf p})$ vanishes only when ${\mathbf x}-{\mathbf x}^\prime$ lies along the $z_e$ axis (see Eq.~(24) [therein], and discussion of this integral in Ref.~\cite{AhluwaliaGRU})
	\begin{align}
		{\mathbf x}-{\mathbf x}^\prime \;\mbox{along }z_e:\quad
		& \{\Lambda({\bf x},t),\; \Pi({\bf x}^\prime,t)\} 
		\na
		=  i \delta^3({\bf x} -{\bf x}^\prime) I  .
\label{eq:LPac}
	\end{align}
The anticommutators for the particle/antiparticle annihilation and creation
operators suffice to yield the remaining locality conditions,
	\begin{align}
		\{\Lambda({\bf x},t),\; \Lambda({\bf x}^\prime,t)\} =
O
,\quad \{\Pi({\bf x},t),\; \Pi({\bf x}^\prime,t)\}
		\label{eq:LLPPac} = 
O
.
	\end{align}

The set of anticommutators contained in Eqs.~(\ref{eq:LPac}) and
(\ref{eq:LLPPac}) establish that $\Lambda(x)$ becomes local along the $z_e$ axis. For this reason we call $z_e$ as the dark axis of locality.}"

Next, I cite Rodrigues {\it et al.}, Ref.~\cite{Rodrigues-PR}:
``{\it We have shown through explicitly and detailed calculation that the integral of
$\mathcal{G}(\mathbf{p})$ appearing in Eq. (42) of \cite{Ahlu-PR} \ is null for
$\mathbf{x-x}^{\prime}$ lying in three orthonormal spatial directions in the
rest frame of an arbitrary inertial frame $\mathbf{e}_{0}=\partial/\partial t$.

This shows that the existence of elko spinor fields does not implies in any
breakdown of locality concerning the anticommutator of \linebreak $\{\Lambda
(\mathbf{x,}t),\Pi(\mathbf{x}^{\prime},t\}$ and moreover does not implies in
any preferred spacelike direction field in Minkowski spacetime.}"

Who is right? In 2013 W. Rodrigues~\cite{Rodrigues-IJTP} changed a bit his opinion. He wrote: 
``{\it When $\Delta_{z}\neq0$, $\mathcal{\hat{G}}(\mathbf{x-x}^{\prime})$ is null the
anticommutator is \emph{local} and thus there exists in the elko theory\ as
constructed in \cite{Ahlu-PR} an infinity number of
\emph{\textquotedblleft}locality\ directions\emph{\textquotedblright.} On the
other hand $\mathcal{\hat{G}}(\mathbf{x-x}^{\prime})$ is\ a distribution with
support in $\Delta_{z}=0$. So$,$ the directions $\mathbf{\Delta}=(\Delta
_{x},\Delta_{y},0)$\ are nonlocal in each arbitrary inertial reference frame
$\mathbf{e}_{0}$ chosen to evaluate $\mathcal{\hat{G}}(\mathbf{x-x}^{\prime}%
)$}", thus accepting the Ahluwalia et al. viewpoint.

Meanwhile, I suggest to use the 8-component (or 16-component) formalism  in similarity with the 12-component formalism of this subsection.  If we calculate\footnote{For $\Psi_\pm^\sigma$ see Eq. (33).}
\begin{align}
 S_F^{(+,-)} (x_2, x_1) &{=} \hspace{-5pt}\int\hspace{-5pt} \frac{d^3 p}{(2\pi)^3}\frac{m}{E_p} \bigg [
\theta (t_2 {-} t_1) \, a \,\,
\Psi_\pm^\sigma (p)  \overline \Psi_\pm^{\sigma} (p) e^{-ip\cdot x} 
\na
  + \theta (t_1 - t_2) \, b \,\,
\Psi_\mp^\sigma (p) \overline \Psi_\mp^\sigma (p) e^{ip\cdot x} \bigg]
\nonumber\\
&=\int 
\frac{d^4 p}{(2\pi)^4} e^{-ip\cdot x} \frac{(\hat p\pm m)}{p^2 -m^2
+i\epsilon}, 
\end{align}
we easily come to the result that the corresponding Feynman-Dyson propagator gives  the local theory 
in the sense:
\begin{align}
\sum_{\pm} [ i\Gamma_\mu \partial^\mu_2 \mp m] S_F^{(+,-)} (x_2 - x_1) = \delta^{(4)} (x_2 - x_1).
\end{align}
However, physics should choose only one correct formalism. It is not clear, why two correct mathematical formalisms  lead to different physical results? First of all, we should check, whether this possible non-locality in the propagators has influence on the physical observables such as the scattering amplitudes, the energy spectra and the decay widths. If not, we may find some unexpected symmetries in relativistic quantum mechanics/field theory. This is the task for future publications. However, it is already obvious
if we would not enlarge the number of components in the fields (in the propagator) we would not be able to obtain the formally causal propagators for higher spins and/or for the neutral particles.

The dilemma of the (non)local propagators for the spin $S=1$ has also been analized 
in~\cite{Kruglov2}
within the Duffin-Kemmer-Petiau (DKP) formalism or the Dirac-K\"ahler formalism~\cite{Kruglov1}. However, the propagators given 
in~\cite{Kruglov2} are those in the generalized Duffin-Kemmer-Petiau formalism, in fact. They are 
not in the Weinberg-Tucker-Hammer  formalism. Moreover, the problem of the massless limit
was not discussed in the DKP formalism, which is non-trivial (like that of the Proca formalism~\cite{Stepan}).

We should use the obtained set of  Weinberg propagators
(\ref{propa1}-\ref{propa4})
in the perturbation calculus of scattering amplitudes.
In Ref.~\cite{Dvoegl-IJTP} the amplitude for the interaction
of two $2(2S+1)$ bosons has been obtained on the basis
of the use of one field only and it is obviously incomplete,
see also Ref.~\cite{Hammer}. But, it is interesting to note
that the spin structure was proved there
to be the same,  regardless we consider
the two-Dirac-fermion interaction or the two-Weinberg $S=1$-boson
interaction. However, the denominator slightly differs ($1/\vec \Delta^2
\rightarrow 1/2m(\Delta_0 -m)$) in the cited papers~\cite{Dvoegl-IJTP}
from the fermion-fermion case, where $(\Delta_0, \vec \Delta )$ is the 
momentum-transfer 4-vector in the Lobachevsky space. 
More accurate considerations
of the fermion-boson and boson-boson interactions in the framework
of the Weinberg theory has been reported elsewhere~\cite{Dvoegl-Valladolid}.
So, the conclusion of this subsection is: one can construct analogs of the Feynman-Dyson
propagators for the $2(2S+1)$ model and, hence, local theories provided that the Weinberg states are
quadrupled  ($S=1$ case), and the neutral particle states are doubled.

\section{Negative-energy and Tachyonic Solutions.}

What is the physical sense of the presented mathematical formalism?
Why did we consider four field functions in the propagator for spin-1 in Ref.~\cite{DvoeglazovBW,Dvoe-RMF}?
Let us make some additional observations for spin-1/2 and spin 1. 

We have 4 solutions in the original Dirac equation for $u-$ and 4 solutions  for 
$v=\gamma^5 u$  (remember we have $p_0=\pm E_p =
\pm \sqrt{{\bf p}^2 +m^2}$). See, for example, Ref.~\cite{Dvoe3}. In the $S=1$ Weinberg equation~\cite{Weinberg} we have 12 solutions.\footnote{In Ref.~\cite{Hammer} we have  causal solutions only for the S=1 Tucker-Hammer equation.} Apart $p_0=\pm E_p$ we have taquionic
solutions $p_0=\pm E'_p =\pm \sqrt{{\bf p}^2 -m^2}$, {\it i. e.} $m\rightarrow im$.  This is easily to be checked on using
the algebraic equations and solving them with respect to $p_0$: 
\begin{equation}
Det [\gamma^{\mu}p_\mu  \pm m ]  =0\,,
\end{equation}
and
\begin{equation}
Det [\gamma^{\mu\nu}p_\mu p_\nu \pm m^2 ]  =0\,,
\end{equation}
where $\gamma^{\mu\nu}$ are again the Barut-Muzinich-Williams $6\times 6$ covariantly-defined 
matrices~\cite{Barut,Weinberg}.
In construction of field operator
we generally need $u(-p)= u (-p_0, -{\bf p}, m)$ which should be transformed to $v (p)=\gamma^5 u (p) = \gamma^5 u (+p_0, +{\bf p}, m)$. On the other hand, when we calculate the parity properties we need ${\bf p} \rightarrow  -{\bf p}$. The $u(p_0, -{\bf p}, m)$ satisfies
\begin{equation}
[\tilde\gamma^{\mu\nu}p_\mu p_\nu -m^2 ] u (p_0, -{\bf p}, m) =0\,.
\end{equation}
The $u (-p_0, {\bf p}, m)$ ``spinor" satisfies: 
\begin{equation}
[\tilde\gamma^{\mu\nu}p_\mu p_\nu -m^2 ] u (-p_0, +{\bf p}, m) =0\,,
\end{equation}
that is the same as above. The tilde signifies $\tilde\gamma^{\mu\nu}= \gamma_{00}\gamma^{\mu\nu}\gamma_{00}$ that is analogoues to the $S=1/2$ case $\tilde\gamma^{\mu}= \gamma_{0}\gamma^{\mu}\gamma_{0}$.
The $u (-p_0, -{\bf p}, m)$ satisfies:  
\begin{equation}
[\gamma^{\mu\nu}p_\mu p_\nu -m^2 ] u (-p_0, -{\bf p}, m) =0\,.
\end{equation}

This case is different from the spin-1/2 case where the spinor $u (-p_0, {\bf p}, m)$
satisfies
\begin{equation}
[\tilde\gamma^{\mu}p_\mu  +m ] u (-p_0, +{\bf p}, m) =0\,,
\end{equation}
and $u (p_0, -{\bf p}, m)$,
\begin{equation}
[\tilde\gamma^{\mu}p_\mu  -m ] u (p_0, -{\bf p}, m) =0\,,
\end{equation}
and 
\begin{equation}
[\gamma^{\mu}p_\mu  +m ] u (-p_0, -{\bf p}, m) =0\,.
\end{equation}

In general we can use  either $u (-p_0, +{\bf p}, m)$ or $u (p_0, -{\bf p}, m)$ to construct the causal propagator
in the spin-1/2 case. However,  a) $u (-p_0, +{\bf p}, m)$ satisfies the similar equation as 
$u (+p_0, -{\bf p}, m)$ and b) we have the integration over ${\bf p}$. This integration is invariant with respect to
${\bf p} \rightarrow -{\bf p}$:
\begin{eqnarray} 
&&S_F (x_2, x_1) =\sum_\sigma\int \frac{d^3 {\bf p}}{(2\pi)^3}\frac{m}{E_p}\\
&& \hspace{-10mm} [
\theta (t_2 -t_1) \, a \,\,
u^\sigma (p)  \overline u^{\sigma} (p) e^{-i{\bf p}\cdot ({\bf x_2 -x_1}) }
+  
\theta (t_1 - t_2) \, b \,\,
v^\sigma (p)  \overline v^\sigma (p) e^{+i{\bf p}\cdot ({\bf x_2 -x_1}) } ] \, .\nonumber
\end{eqnarray}
So the results for the causal propagator ($S=1/2$)
\begin{equation}\label{dp2}
S_F (x) = \int \frac{d^4 p}{(2\pi)^4} e^{-ip\cdot x} \frac{\hat p \pm m}{p^2 -m^2
+i\epsilon} \,
\end{equation}
do not change physics  observables regardless the use of one or another Dirac equation.

The situation is somewhat different for spin 1. 
The tachyonic solutions of the original Weinberg  equation
\begin{equation}
[\gamma^{\mu\nu}p_\mu p_\nu -m^2 ] u (p_0, +{\bf p}, m) =0\label{w11}
\end{equation}
are just the solutions 
of the equation with the opposite square of $m \rightarrow im$):
\begin{equation}
[ \gamma^{\mu\nu}p_\mu p_\nu +m^2 ] u (p_0, +{\bf p}, im) =0\,.\label{w12}
\end{equation}
We cannot transform the propagator of the original equation (\ref{w11}) to that of (\ref{w12}) just by the change of the variables
as in the spin-1/2 case. The mass square changed the sign, just as in the case of $v-$ ``spinors". When we construct 
the propagator 
we have to take into account this solution and the superposition $u (p, m)$ and $u (p, im)$, and corresponding equations. 

The conclusion is paradoxical: in order to construct the causal propagator for the  spin 1
we have to take acausal (tachyonic) solutions of homogeneous equations into account. It is not surprising that the propagator is not causal for the Tucker-Hammer equation because it does not contain the tachyonic solutions. Probably, this statement  is valid for all higher spins.


\section{The Conclusions.}       

First of all, the point of my paper is: there are ``negative-energy solutions" in that is previously  considered as 
``positive-energy solutions" of relativistic wave equations, and vice versa. Their explicit forms have been presented. For example in the $S=1/2$ case both algebraic equation  $Det (\hat p - m) =0$ and $Det (\hat p + m) =0$  for  $u-$ and $v-$  4-spinors have solutions with $p_0= \pm E_p =\pm \sqrt{{\bf p}^2 +m^2}$. The same is true for higher-spin equations. Meanwhile, every book  applies the Dirac-Feynman-Stueckelberg procedure for elimination of negative-energy solutions. The recent Ziino works (and, independently, 
the articles of several other authors) show that the Fock space can be doubled. We re-consider this possibility on the quantum-field level for both  
$S=1/2$ and higher spin particles. Next, the relations to the previous works have been found. For instance,
the doubling of the Fock space and the corresponding solutions of the Dirac equation, and those of higher-spin equations
obtained  additional mathematical bases in this paper. The tachyonic solutions seem to play crucial role in construction of the propagators for higher spins.
.


\section*{Acknowledgments.}  

I acknowledge discussions with Prof. N. Debergh, the late Prof. W. Rodrigues, Jr. and the late Prof. Z. Oziewicz.
I am grateful to the Zacatecas University for professorship.
I appreciate discussions with participants of several recent Conferences, and referee recommendations.


}

\end{document}